\documentclass[aps,twocolumn,prl,superscriptaddress,preprintnumbers,showpacs,tightenlines]{revtex4}
\usepackage{amssymb}
\usepackage{times}
\usepackage{graphicx}
\usepackage{epsfig}
\usepackage{color}
\usepackage{multirow}
\usepackage{threeparttable}
\usepackage{bm}

\begin{document}
\title{Theoretical spin-wave dispersions in the antiferromagnetic phase AF1 of MnWO$_4$ based on the polar atomistic model in \emph{P}2}
\author{B.-Q. Liu}
\affiliation{Key Laboratory of Neutron Physics, Institute of Nuclear Physics and Chemistry, CAEP, Mianyang 621900, PR China}
\affiliation{J\"ulich Centre for Neutron Science (JCNS) at Heinz Maier-Leibnitz Zentrum (MLZ), Forschungszentrum J\"ulich GmbH, Lichtenbergstrasse 1, 85748 Garching, Germany}
\author{S.-H. Park}
\affiliation{Section Crystallography, Department for Earth and Environmental Sciences, Ludwig-Maximilians-Universitaet Muenchen, Theresienstrasse 41, 80333 Munich, Germany}
\author{P. \v Cerm\'ak}
\author{A. Schneidewind}
\affiliation{J\"ulich Centre for Neutron Science (JCNS) at Heinz Maier-Leibnitz Zentrum (MLZ), Forschungszentrum J\"ulich GmbH, Lichtenbergstrasse 1, 85748 Garching, Germany}
\author{Y. Xiao}
\email{xiaoyg@pkusz.edu.cn}
\affiliation{School of Advanced Materials, Peking University, Peking University Shenzhen Graduate School, Shenzhen 518055, PR China}


\begin{abstract}
The spin wave dispersions of the low temperature antiferromagnetic phase (AF1) MnWO$_4$ have been numerically calculated based on the recently reported non-collinear spin configuration with two different canting angles. A Heisenberg model with competing magnetic exchange couplings and single-ion anisotropy terms could properly describe the spin wave excitations, including the newly observed low-lying energy excitation mode $\omega_2$=0.45 meV appearing at the magnetic zone centre. The spin wave dispersion and intensities are highly sensitive to two differently aligned spin-canting sublattices in the AF1 model. Thus this study reinsures the otherwise hardly provable hidden polar character in MnWO$_4$.

\end{abstract}

\maketitle

\section{I. INTRODUCTION}

Multiferroic properties, which show ferroelectricity or ferroelasticity coexisting with magnetic order, have attracted great attention both experimentally and theoretically. Of particular interest for such materials is that they may have potential applications in electronic devices like magnetoelectric sensors and data storage chips \cite{M.Maczka2012}. A number of materials such as $R$MnO$_3$ ($R$ is rare earth element) \cite{T.Kimura2003,T.Goto2004}, $R$Mn$_2$O$_5$ \cite{N.Hur2004,D.Higashiyama2005}, CoCr$_2$O$_4$ \cite{Y.Yamasaki2006} exhibiting a strong interplay between the magnetic and ferroelectric order have been intensively studied. Several different models have been proposed to explain the mechanism of magnetoelectric effects \cite{H.Katsura2005,I.A.Sergienko2006,M.Mostovoy2006,A.B.Harris2007,T.Arima2007}. For examples, the change of the modulation wavelength seems to play an important role \cite{L.C.Chapon2006}, another key factor can be a noncollinear spin configuration \cite{M.Kenzelmann2005,T.Arima2006} which is in accord with the theory associated with the Aharonov-Casher effect \cite{Y.Aharonov1984} or the inverse Dzyaloshinskii-Moriya (DM) interaction \cite{I.A.Sergienko2006}.

Another well-known material is the mineral huebnerite MnWO$_4$, an exemplary prototype of magnetoelectric control. It is also a promising system for the study of magnetic phase transitions and related critical phenomena, since it exhibits rich magnetic phase diagram by chemical substitutions \cite{K.Taniguchi2006,A.H.Arkenbout2006} and by applying magnetic fields \cite{H.Mitamura2012}. At zero field, three successive antiferromagnetic phase transitions are observed: the commensurate magnetic structure AF1 with propagation vector $\mathbf{k}=(\pm\frac{1}{4},\frac{1}{2},\frac{1}{2})$ is present below 8 K, the incommensurate elliptical spiral spin structure AF2 existing in 8 $\sim$ 12.3 K induces ferroelectric order, and the incommensurate collinear sinusoidal spin structure AF3 is only observed in a narrow temperature range below 13.5 K \cite{A.H.Arkenbout2006,H.Dachs1969,G.Lautenschlaeger1993}. The fundamental crystal structure of MnWO$_4$ is monoclinic, and the corresponding space group has been believed to be $\emph{\textbf{P}}2/c$ until our structural studies confirmed the true symmetry $\emph{\textbf{P}}2$ \cite{S.H.Park2015,U.Gattermann2016-1,U.Gattermann2016-2,S.H.Park2018} and the two different spin-canting configurations at two Mn$^{2+}$ sublattices, as a symmetrical consequence of the direct polar subgroup relation between $\emph{\textbf{P}}2$ and $\emph{\textbf{P}}2/c$. This non-collinear spin-canting texture in AF1 of MnWO$_4$ is in contrast to the previous collinear magnetic model ($\uparrow\uparrow\downarrow\downarrow$) where the magnetic moments at two Mn sites are aligned collinearly along the easy axis with a common angle of 35$^\circ$-37$^\circ$ against $a$-axis on the (\textit{a-c}) plane \cite{A.H.Arkenbout2006,H.Dachs1969,G.Lautenschlaeger1993}. With this non-collinear magnetic model, it is necessary to re-examine the excitation spectra as well as the corresponding exchange coupling interactions in the AF1 MnWO$_4$ as they are sensitive to the spin configurations.

It is known that MnWO$_4$ is a frustrated magnet with competing exchange interactions \cite{G.Lautenschlaeger1993,H.Ehrenberg1999}. The analysis of the magnetic excitations allows to explore and clarify the underlying magnetic interactions which dominate the complex spin configurations. There have been several theoretical and experimental investigations \cite{H.Ehrenberg1999,H.Ehrenberg1996,Ye2011,Xiao2016,Tian2009} on spin wave dispersions in AF1 based on the centrosymmetric space group \emph{\textbf{P}}2/$c$, where 9$\sim$12 exchange coupling parameters as well as single-ion anisotropy parameter are evaluated within a Heisenberg model. Among these work, in the recent high-resolution inelastic neutron scattering (INS) study of AF1 phase MnWO$_4$, Xiao \textit{et al.} \cite{Xiao2016} revealed two new electromagnon branches appearing at low energies of $\omega_1$=0.07 meV and $\omega_2$=0.45 meV at the zone centre, which may reflect the dynamical magnetoelectric coupling and cannot be described by the Heisenberg model.

In this work, we present spin wave calculations based on the polar structure of MnWO$_4$ (space group $\emph{\textbf{P}}2$) as well as the non-collinear magnetic structure, demonstrating that the spin wave dispersion in AF1 can be described by a Heisenberg model with 11 magnetic exchange coupling parameters and single-ion anisotropy. The calculated spectra are visualized in a proper way for easy comparison with previous experimental INS data. Interestingly, one of the electromagnon excitation modes previously denoted as $\omega_2$ may be properly described in this study.

\section{II. Theoretical calculations of spin wave dispersion}

As aforementioned, the most recent study \cite{S.H.Park2018} showed that MnWO$_4$ crystallises in monoclinic $\emph{\textbf{P}}2$ structure, and the low temperature magnetic structure is not a collinear spin configuration but two spin-canting textures, as illustrated in Fig. 1. The magnetic spins lie in ($a$-$c$) plane with the Mn$_a$ spin-canted about $\theta_1=33^\circ$ from the $a$-axis while Mn$_b$ about $\theta_2=59^\circ$.

\begin{figure}
\centering
\includegraphics[width=8.5cm]{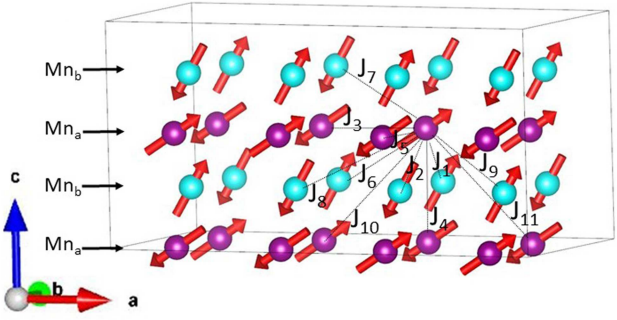}
\hspace{0.5cm} \caption{(Color online) The magnetic structure of the AF1 phase of MnWO$_4$, showing the magnetic Mn$^{2+}$ ions only. Two different spin-canting textures are indicated by the directions of magnetic moments (arrows) of Mn$^{2+}$ ions at both unique sites at Mn$_a$ (in pink) and Mn$_b$ (in cyan blue). Eleven exchange coupling constants $J_1$ to $J_{11}$ (dashed lines) are used to fit spin wave dispersions in this study.}
\end{figure}

The spin wave dispersion curves can be modeled by an effective Heisenberg Hamiltonian
\begin{equation}
H=-\frac{1}{2}\sum_{i,l}{J_{i,l}\mathbf{S}_i \cdot \mathbf{S}_l}-D_a\sum_i{(\mathbf{a}\cdot \mathbf{S}_i)^2}-D_c\sum_i{(\mathbf{c}\cdot \mathbf{S}_i)^2},
\end{equation}
where $J_{i,l}$ denotes exchange coupling constants (from $J_1$ to $J_{11}$), $D_{a(c)}$ denotes single-ion anisotropy constant.

It should be noted that, for each point occupied by a magnetic atom, an individual axis of quantization is introduced, and with each point $i(l)$ associate a local coordinate system ($x$, $y$, $z$) so that the $z$ axis in this system coincides with the equilibrium spin direction at this point \cite{Y.A.Izyumov1970}. For example, the transformation of the vector $\mathbf{S}_1$ from the general system of coordinates ($x'$, $y'$, $z'$) associated with the crystallographic axes to the local coordinate system is
\begin{eqnarray}
S_1^{x'}=S_1^x\sin{\theta_1}+S_1^z\cos{\theta_1},\nonumber\\
S_1^{y'}=S_1^y,\nonumber\\
S_1^{z'}=S_1^z\sin{\theta_1}-S_1^x\cos{\theta_1}.\nonumber
\end{eqnarray}

Introducing the notation $i$ for spin up ($\uparrow$) sites and $l$ for spin down ($\downarrow$) sites, the linearized Holstein-Primakoff transformation for the quantum spin $\mathbf{S}$ at each site \cite{T.Holstein1940} can be written as
\begin{equation}
S_{i,l}^\dagger=S_{i,l}^x+iS_{i,l}^y,  S_{i,l}^-=S_{i,l}^x-iS_{i,l}^y,
\end{equation}
and
\begin{eqnarray}
S_i^\dagger=\sqrt{2S}\bigg(1-\frac{a_i^\dagger a_i}{2S}\bigg)^{1/2}a_i\approx\sqrt{2S}a_i,\\
S_i^-=\sqrt{2S}\bigg(1-\frac{a_i^\dagger a_i}{2S}\bigg)^{1/2}a_i^\dagger\approx\sqrt{2S}a_i^\dagger, \\
S_i^z=S-a_i^\dagger a_i, \\
S_l^\dagger=\sqrt{2S}\bigg(1-\frac{b_l^\dagger b_l}{2S}\bigg)^{1/2}b_l^\dagger\approx\sqrt{2S}b_l^\dagger,\\
S_l^-=\sqrt{2S}\bigg(1-\frac{b_l^\dagger b_l}{2S}\bigg)^{1/2}b_l\approx\sqrt{2S}b_l,\\
S_l^z=-S+b_l^\dagger b_l.
\end{eqnarray}

The Fourier transformation is introduced by:
\begin{eqnarray}
a_i^\dagger=\sqrt{\frac{1}{N}}\sum_\mathbf{q}{\exp{(-i\mathbf{q}\cdot \mathbf{r}_i)}a_\mathbf{q}^\dagger},\\
a_i=\sqrt{\frac{1}{N}}\sum_\mathbf{q}{\exp{(i\mathbf{q}\cdot \mathbf{r}_i)}a_\mathbf{q}}, \\
b_l^\dagger=\sqrt{\frac{1}{N}}\sum_\mathbf{q}{\exp{(-i\mathbf{q}\cdot \mathbf{r}_l)}b_\mathbf{q}^\dagger}, \\
b_l=\sqrt{\frac{1}{N}}\sum_\mathbf{q}{\exp{(i\mathbf{q}\cdot \mathbf{r}_l)}b_\mathbf{q}}.
\end{eqnarray}

Thus, one can obtain the bosonic Hamiltonian in the momentum space as
\begin{equation}
H=-\frac{1}{2}\sum_q{\mathbf{a}^\dagger H_q\mathbf{a}},
\end{equation}
with $\mathbf{a}^\dagger$=$[a_{q1}^\dagger$, $a_{q2}^\dagger$, $a_{q3}^\dagger$, $a_{q4}^\dagger$, $a_{-q1}$, $a_{-q2}$, $a_{-q3}$, $a_{-q4}$, $b_{q1}^\dagger$, $b_{q2}^\dagger$, $b_{q3}^\dagger$, $b_{q4}^\dagger$, $b_{-q1}$, $b_{-q2}$, $b_{-q3}$, $b_{-q4}]$, and
\begin{equation}
H_q=
\left(
   \begin{array}{cccc}
    M_{11}    & M_{12}  & M_{13}   & M_{14} \\
    M_{12}    & M_{11}  & M_{14}   & M_{13} \\
    M_{13}    & M_{14}  & M_{11}   & M_{12} \\
    M_{14}    & M_{13}  & M_{12}   & M_{11}
   \end{array}
\right),
\end{equation}

\begin{equation}
M_{11}=
\left(
  \begin{array}{cccc}
    A_1     & BP_2   & C       & HP_2    \\
    B^*P_2  & A_2    & H^*P_2  & C^*     \\
    C^*     & HP_2   & A_1     & FP_2    \\
    H^*P_2  & C      & F^*P_2  & A_2
  \end{array}
\right),
\end{equation}

\begin{equation}
M_{12}=
\left(
  \begin{array}{cccccccc}
     A_3       & BP_1   & 0         & HP_1\\
     B^*P_1    & A_4    & H^*P_1    & 0 \\
     0         & HP_1   & A_3       & FP_1 \\
     H^*P_1    & 0      & F^*P_1    & A_4
  \end{array}
\right),
\end{equation}

\begin{equation}
M_{13}=
\left(
  \begin{array}{cccc}
    0         & FP_1   & 0         & GP_1    \\
    F^*P_1    & 0      & G^*P_1    & 0       \\
    0         & GP_1   & 0         & BP_1    \\
    G^*P_1    & 0      & B^*P_1    & 0
  \end{array}
\right),
\end{equation}

\begin{equation}
M_{14}=
\left(
  \begin{array}{cccc}
     E       & FP_2   & C^*     & GP_2 \\
     F^*P_2  & E      & G^*P_2  & C \\
     C       & GP_2   & E       & BP_2 \\
     G^*P_2  & C^*    & B^*P_2  & E
  \end{array}
\right),
\end{equation}

with
\begin{eqnarray}
A_1=&2(S_1J_4+S_1J_5+S_2J_6-S_2J_7-S_2J_8 \nonumber\\
    &+S_2J_9-S_1D_c\sin^2{\theta_1}-S_1D_a\cos^2{\theta_1})\nonumber\\
    &+S_1D_c\cos^2{\theta_1}+S_1D_a\sin^2{\theta_1},
\end{eqnarray}
\begin{eqnarray}
A_2=&2(S_2J_4+S_2J_5+S_1J_6-S_1J_7-S_1J_8  \nonumber \\
    &+S_1J_9-S_2D_c\sin^2{\theta_2}-S_2D_a\cos^2{\theta_2}) \nonumber\\
    &+S_2D_c\cos^2{\theta_2}+S_2D_a\sin^2{\theta_2},
\end{eqnarray}

\begin{equation}
A_3=\frac{1}{2}S_1D_c\cos^2{\theta_1}+\frac{1}{2}S_1D_a\sin^2{\theta_1},
\end{equation}
\begin{equation}
A_4=\frac{1}{2}S_2D_c\cos^2{\theta_2}+\frac{1}{2}S_2D_a\sin^2{\theta_1},
\end{equation}
\begin{eqnarray}
B=&J_1e^{-2\pi i[2(0.41029-0.59519)q_y+0.5q_z]} \nonumber\\
   &+J_2e^{2\pi i[2(0.09519-0.41029)q_y+0.5q_z]},
\end{eqnarray}
\begin{eqnarray}
C=&S_1\big\{J_3e^{-2\pi iq_x}+J_{10}e^{-2\pi i(-q_x-q_z)} \nonumber\\
  &+J_{11}e^{-2\pi i(-q_x+q_z)}\big\},
\end{eqnarray}
\begin{eqnarray}
H=&J_7\big\{e^{-2\pi i[-q_x+2(0.41029-0.59519)q_y+0.5q_z]} \nonumber\\
  &+e^{-2\pi i[q_x+2(0.41029-0.59519)q_y-0.5q_z]}\big\} \nonumber\\
  &+J_8\big\{ e^{-2\pi i[-q_x+2(0.41029-0.09519)q_y-0.5q_z]}  \nonumber\\
  &+e^{-2\pi i[q_x+2(0.41029-0.09519)q_y+0.5q_z]} \big\},
\end{eqnarray}
\begin{equation}
E=S_1[2J_4\cos{q_z}+2J_5\cos{(-q_y)}],
\end{equation}
\begin{eqnarray}
F=&J_2e^{-2\pi i[2(0.41029-0.09519)q_y+0.5q_z]} \nonumber\\
  &+J_1e^{2\pi i[2(0.59519-0.41029)q_y+0.5q_z]},
\end{eqnarray}
\begin{eqnarray}
G=&J_6\big\{e^{2\pi i[-q_x+2(0.59519-0.41029)q_y-0.5q_z]} \nonumber\\
  &+e^{2\pi i[q_x+2(0.59519-0.41029)q_y+0.5q_z]} \big\} \nonumber\\
  &+J_9\big\{ e^{2\pi i[-q_x+2(0.41029-0.59519)q_y+0.5q_z]} \nonumber\\
  &+e^{2\pi i[q_x+2(0.41029-0.59519)q_y-0.5q_z]} \big\},
\end{eqnarray}
\begin{equation}
P_{1,2}=\frac{1}{2}\sqrt{S_1S_2}[\cos{(\theta_1-\theta_2)}\mp1].
\end{equation}

In order to diagonalize the Hamiltonian Eq.(\textbf{13}), it is needed to introduce a transformation matrix $\mathbf{T}$ \cite{R.M.White1965}, and $\mathbf{a}=\mathbf{T}\mathbf{\alpha}$ with $\alpha^\dagger$=$[\alpha_{q1}^\dagger$, $\alpha_{q2}^\dagger$, $\alpha_{q3}^\dagger$, $\alpha_{q4}^\dagger$, $\alpha_{-q1}$, $\alpha_{-q2}$, $\alpha_{-q3}$, $\alpha_{-q4}$, $\beta_{q1}^\dagger$, $\beta_{q2}^\dagger$, $\beta_{q3}^\dagger$, $\beta_{q4}^\dagger$, $\beta_{-q1}$, $\beta_{-q2}$, $\beta_{-q3}$, $\beta_{-q4}]$, so that
\begin{equation}
\mathbf{a}^\dagger H_q\mathbf{a}=\mathbf{\alpha}^\dagger \mathbf{T}^\dagger H_q\mathbf{T}\mathbf{\alpha}=\mathbf{\alpha}^\dagger\omega\mathbf{\alpha},
\end{equation}
where $\mathbf{\omega}$ is a diagonal matrix and its diagonal elements are eigenvalues of the system. According to Eq.(\textbf{30}), we obtain $H_q\mathbf{T}=(\mathbf{T}^\dagger)^{-1}\omega$. By using the commutator matrix $\mathbf{I}_1=diag{\{1,1,1,1,-1,-1,-1,-1,1,1,1,1,-1,-1,-1,-1\}}$, one can numerically obtain the system's eigenvalues, i.e., the spin wave excitation energies. The transformation matrix $\mathbf{T}$ is a $16\times16$ matrix with columns that are eigenvectors to $\mathbf{I}_1H_q\mathbf{T}=\omega\mathbf{T}$, and it also must respect the Bose commutation rules. Once the correct transformation matrix $\mathbf{T}$ is obtained, the differential scattering cross sections for magnetic scattering are calculated \cite{S.W.Lovesey1984,T.B.S.Jensen,T.B.S.Jensen2009}, as follows:
\begin{eqnarray}
\frac{d^2\sigma}{d\Omega dE}=\frac{k_f}{k_i}(\gamma r_0)^2\big[\frac{g}{2}F(\mathbf{Q})\big]^2e^{-2W} \nonumber\\
                              \times\sum_{\mu\nu}(\delta_{\mu,\nu}-\hat{Q}_\mu\hat{Q}_\nu)S_{\mu\nu}(\mathbf{Q},\omega),
\end{eqnarray}
with
\begin{equation}
S_{\mu\nu}(\mathbf{Q},\omega)=\frac{1}{2\pi}\int{dte^{-i\omega t}S_{\mu\nu}(\mathbf{Q},t)},
\end{equation}
and
\begin{equation}
S_{\mu\nu}(\mathbf{Q},t)=\sum_{\mathbf{r},\mathbf{r}'}{e^{i\mathbf{Q}\cdot(\mathbf{r}-\mathbf{r}')}\langle S_\mathbf{r}^\mu(0)S_{\mathbf{r}'}^\nu(t)\rangle},
\end{equation}
where $k_f$ and $k_i$ are final and incident wave vectors, respectively; the incident wavelength 4.4 \AA\, was applied to compare INS spectra reported in \cite{Ye2011}; $\gamma r_0$ is the magnetic scattering amplitude for an electron; $g$ is the Land$\acute{\rm e}$ splitting factor for Mn$^{2+}$; $F(\mathbf{Q})$ is dimensionless magnetic form factor; $e^{-2W}$ is the Debye-Waller factor. $\hat{Q}_{\mu(\nu)}$ corresponds to the $\mu(\nu)$ component of a unit vector in the direction of $\mathbf{Q}$. For INS, only the transverse correlations $\langle S_\textbf{r}^x(0)S_{\textbf{r}'}^x(t)\rangle$, $\langle S_\textbf{r}^x(0)S_{\textbf{r}'}^y(t)\rangle$, $\langle S_\textbf{r}^y(0)S_{\textbf{r}'}^x(t)\rangle$, and $\langle S_\textbf{r}^y(0)S_{\textbf{r}'}^y(t)\rangle$ contribute to the cross section, e.g., $\langle S_\textbf{r}^x(0)S_{\textbf{r}'}^x(t)\rangle$ can be written as
\begin{eqnarray}
\langle S_\textbf{r}^x(0)S_{\textbf{r}'}^x(t)\rangle=\frac{1}{4}\bigg[\langle S_\textbf{r}^+(0)S_{\textbf{r}'}^+(t)\rangle+\langle S_\textbf{r}^+(0)S_{\textbf{r}'}^-(t)\rangle \nonumber \\
+\langle S_\textbf{r}^-(0)S_{\textbf{r}'}^+(t)\rangle+\langle S_\textbf{r}^-(0)S_{\textbf{r}'}^-(t)\rangle\bigg].
\end{eqnarray}
Let's calculate the scattering cross section for creating an $\alpha$ spin-wave excitation. For spin up ($\uparrow$) ions $i$ and $i'$, we start with
\begin{eqnarray}
\langle S_i^+(0)S_{i'}^-(t)\rangle = 2S_1\langle a_i(0)a_{i'}^\dagger(t)\rangle\nonumber \\
=2S_1\frac{1}{N}\sum_{\mathbf{q},\mathbf{q}'}{e^{i\mathbf{q}\cdot \mathbf{r}_i}e^{-i\mathbf{q}'\cdot \mathbf{r}_{i'}}\langle a_\mathbf{q}(0)a_{\mathbf{q}'}^\dagger(t)\rangle}.
\end{eqnarray}

Only the part of $\langle a_\mathbf{q}(0)a_{\mathbf{q}'}^\dagger(t)\rangle$ describes the creation of an $\alpha$ spin-wave excitation, i.e., transforms into $\langle\alpha_\mathbf{q}(0)\alpha_\mathbf{q}^\dagger(t)\rangle=\big(n(\omega_\mathbf{q}^\alpha)+1\big)e^{i\omega_\mathbf{q}^\alpha t}$, with $n(\omega)=(e^{\frac{\hbar\omega}{k_BT}}-1)^{-1}$. Therefore, by using the relationship between $a_\mathbf{q}$ and $\alpha_\mathbf{q}$, as well as the delta function $\delta_{\mathbf{q},\mathbf{q}'}$, the intensity of spin-wave excitation can finally be calculated for the respective configurations, e.g.,
\begin{eqnarray}
S_{\uparrow\uparrow}^{+-}(\mathbf{Q},\omega_1)^{\alpha+}=\sum_{\tau}{F_{1\uparrow}(\tau)F_{1\uparrow}^*(\tau)T_{1,1}T_{1,1}^*n^+(\omega_1^\alpha)}\nonumber\\
+\sum_{\tau}{F_{1\uparrow}(\tau)F_{2\uparrow}^*(\tau)T_{1,1}T_{2,1}^*n^+(\omega_1^\alpha)}\nonumber\\
+\sum_{\tau}{F_{1\uparrow}(\tau)F_{3\uparrow}^*(\tau)T_{1,1}T_{3,1}^*n^+(\omega_1^\alpha)}\nonumber\\
+\sum_{\tau}{F_{1\uparrow}(\tau)F_{4\uparrow}^*(\tau)T_{1,1}T_{4,1}^*n^+(\omega_1^\alpha)},
\end{eqnarray}
where the indices of $S_{\uparrow\uparrow}^{+-} (\textbf{Q},\omega_1)^{\alpha+}$ mean that we are considering the contribution to the cross section of one branch (denoted by $\omega_1$) for creating an $\alpha$ spin-wave excitation (denoted by $^{\alpha+}$) from the thermal mean value of $\langle S_i^+(0)S_{i'}^-(t)\rangle$ in Eq.(35). As $i$ and $i'$ define spin up, we write $_{\uparrow\uparrow}$, and for dealing with two operators $S^+$ and $S^-$ we use $S^{+-}$. In Eq.(36), $\tau$ is reciprocal lattice vector, $T_{i,l}$ are the matrix elements of $\mathbf{T}$, and the sum extends over all spins with the same type in a magnetic unit cell. Considering both direction and magnitude of the magnetic moment, there are four different types of spins as shown in Fig. 1, distinguished by the subscript numbers $\epsilon$ ($\epsilon=$1,2,3,4) in the $F$ factor. $F_\epsilon(\tau)=\sum_\mathbf{d}{e^{-i\tau\cdot \mathbf{d}}}$ is spin dependent structure factor, with $\mathbf{d}$ the position of the magnetic ion with $\epsilon$ type. Similarly, further existing spin-wave excitations could be calculated, as follows:
\begin{eqnarray}
S_{\uparrow\downarrow}^{+-}(\mathbf{Q},\omega_1)^{\alpha+}=\sum_{\tau}{F_{1\uparrow}(\tau)F_{1\downarrow}^*(\tau)T_{1,1}T_{13,1}^*n^+(\omega_1^\alpha)}\nonumber\\
+\sum_{\tau}{F_{1\uparrow}(\tau)F_{2\downarrow}^*(\tau)T_{1,1}T_{14,1}^*n^+(\omega_1^\alpha)}\nonumber\\
+\sum_{\tau}{F_{1\uparrow}(\tau)F_{3\downarrow}^*(\tau)T_{1,1}T_{15,1}^*n^+(\omega_1^\alpha)}\nonumber\\
+\sum_{\tau}{F_{1\uparrow}(\tau)F_{4\downarrow}^*(\tau)T_{1,1}T_{16,1}^*n^+(\omega_1^\alpha)},
\end{eqnarray}
\begin{eqnarray}
S_{\downarrow\uparrow}^{+-}(\mathbf{Q},\omega_1)^{\alpha+}=\sum_{\tau}{F_{1\downarrow}(\tau)F_{1\uparrow}^*(\tau)T_{13,1}T_{1,1}^*n^+(\omega_1^\alpha)}\nonumber\\
+\sum_{\tau}{F_{1\downarrow}(\tau)F_{2\uparrow}^*(\tau)T_{13,1}T_{2,1}^*n^+(\omega_1^\alpha)}\nonumber\\
+\sum_{\tau}{F_{1\downarrow}(\tau)F_{3\uparrow}^*(\tau)T_{13,1}T_{3,1}^*n^+(\omega_1^\alpha)}\nonumber\\
+\sum_{\tau}{F_{1\downarrow}(\tau)F_{4\uparrow}^*(\tau)T_{13,1}T_{4,1}^*n^+(\omega_1^\alpha)},
\end{eqnarray}
\begin{eqnarray}
S_{\downarrow\downarrow}^{+-}(\mathbf{Q},\omega_1)^{\alpha+}=\sum_{\tau}{F_{1\downarrow}(\tau)F_{1\downarrow}^*(\tau)T_{13,1}T_{13,1}^*n^+(\omega_1^\alpha)}\nonumber\\
+\sum_{\tau}{F_{1\downarrow}(\tau)F_{2\downarrow}^*(\tau)T_{13,1}T_{14,1}^*n^+(\omega_1^\alpha)}\nonumber\\
+\sum_{\tau}{F_{1\downarrow}(\tau)F_{3\downarrow}^*(\tau)T_{13,1}T_{15,1}^*n^+(\omega_1^\alpha)}\nonumber\\
+\sum_{\tau}{F_{1\downarrow}(\tau)F_{4\downarrow}^*(\tau)T_{13,1}T_{16,1}^*n^+(\omega_1^\alpha)}.
\end{eqnarray}
The calculation for $S^{++}$, $S^{-+}$, $S^{--}$, as well as the contribution to the scattering cross section for creating a $\beta$ magnon is almost the same. The numerical calculations are performed by self-developed Fortran code.

\section{III. Results and discussion}

Since the spin-canting structure has been employed in this work, there will be 8 branches of spin wave dispersions instead of 4 branches for the collinear model. Fig. 2 shows the spin wave dispersions along [H,0.5,2H] direction through the magnetic Bragg peak (0.25,0.5,0.5). The solid lines denote the spin wave dispersion relationship from a fit of previous experimental data \cite{Xiao2016} by a Heisenberg model as described above. One can find that most of them are degenerated, only some splitting which are resolvable for several branches. Interestingly, the lowest branch at an energy level about 0.45 meV resembles the electromagnon excitation mode $\omega_2$ observed in Ref.\cite{Xiao2016}. As away from the magnetic zone centre, the calculated intensity of this branch will first increases with H and then decreases dramatically near H=0.3. At present, we assume that this $\omega_2$ may be magnons which arise from the spin-canting structure with two different canting angles $\theta_1$ and $\theta_2$. The calculated spin wave spectrum along [H,0.5,2H] with Gaussian function convoluted is shown in Fig. 3, which consistently captures the characters of the experimental spectrum such as the strongly asymmetric intensity around the magnetic zone centre. There is a spin gap of $\sim$ 0.5 meV and boundary energy about 2.2 meV, which are also in good agreement with the previous experimental spectra \cite{Ye2011,Xiao2016}.

\begin{table*}[htbp]
\begin{threeparttable}
\caption{Magnetic exchange coupling constants evaluated from spin wave model calculation are compared with those from previous studies. The distance (in unit of \AA) between two interacting Mn spins are listed for the respective corresponding exchange coupling constants (in unit of meV).}
\begin{tabular}{lllllllllllllll}
\toprule
                         & $J_1$ &  $J_2$ & $J_3$  & $J_4$  & $J_5$  & $J_6$  & $J_7$ & $J_8$  & $J_9$   & $J_{10}$ & $J_{11}$ & $J_{12}$  & $D_c$ & $D_a$\\
\hline
Mn-Mn-distance           & 3.271 &  4.394 & 4.816  & 4.979  & 5.743  & 5.784  & 5.860 & 6.485  & 6.552   &  6.863   & 6.990    & $\cdots$  &     &  \\
This work                & -0.4  & -0.002 & -0.19  &  -0.28 & -0.01 & -0.34  & -0.12 & -0.01  & -0.29   & -0.12    &-0.04    &$\cdots$   & 0.08  & 0.06 \\
Ref.\cite{H.Ehrenberg1999} &-0.084 & -0.058 & -0.182 & 0.178  & 0.009  & -0.219 & 0.01  & 0.212  &  -0.980 &$\cdots$  & $\cdots$ &$\cdots$   &0.061  & $\cdots$\\
Ref.\cite{Ye2011}        &-0.42  & -0.04  & -0.32  &  -0.26 & 0.05   & -0.43  & -0.12 & 0.02   &  -0.26  & -0.15    & 0.02     &$\cdots$   & 0.09  & $\cdots$\\
Ref.\cite{Xiao2016}      &-0.37  &-0.002  &-0.17   & -0.21  &-0.011  & -0.34  &-0.11  &-0.010  &-0.20    &-0.12     &-0.042    &-0.016     &0.06  &$\cdots$ \\
Ref.\cite{Tian2009}      &-0.16  & -0.153 & -0.232 & -0.018 & -0.089 & -0.185 &-0.031 & -0.115 &$\cdots$ &$\cdots$  & $\cdots$ &$\cdots$   &$\cdots$ &$\cdots$ \\
\toprule\\
\end{tabular}
\end{threeparttable}
\end{table*}

\begin{figure}
\centering
\includegraphics[width=8.5cm]{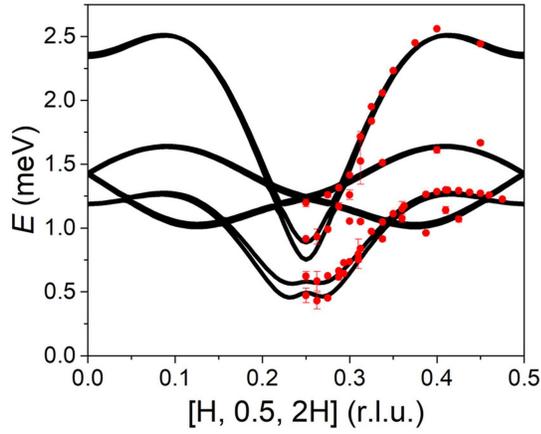}
\hspace{0.5cm} \caption{(Color online) Spin wave dispersion along [H,0.5,2H] direction through the magnetic peak (0.25,0.5,0.5), the experimental data (red points) are taken from Ref.\cite{Xiao2016}, with the lowest branch at $\approx$0.45 meV previously ascribed to electromagnon excitation mode $\omega_2$, fitted by the spin-canting model (solid line).}
\end{figure}

\begin{figure}
\centering
\includegraphics[width=8.5cm]{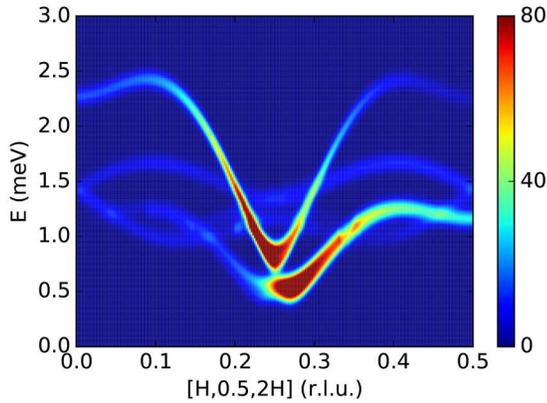}
\hspace{0.5cm} \caption{(Color online) Spin wave spectrum along [H,0.5,2H] direction through the magnetic peak (0.25,0.5,0.5), with Gaussian function convoluted. The color code denotes the INS intensity.}
\end{figure}

Fig. 4 is spin wave dispersion along [0.25,K,0.5]. The fitting results exhibit acceptable agreement with the measured spin wave excitations, with all parameters listed in Table I, along with previous experimental and theoretical studies. The low-lying energy excitations located in the zone centre may still be described by the spin-canting model, although one of the spin wave branches does not match the experimental data perfectly. In Fig.5, the magnetic scattering spectrum calculated in [0.25,K,0.5] crossing the magnetic reflection (0.25,0.5,0.5) also properly describes the observed scattering intensity map (Fig. 4a in Ref. \cite{Ye2011}), where for the lowest branch the measured scattering intensity is strong on both sides of the magnetic zone centre.

\begin{figure}
\centering
\includegraphics[width=8.5cm]{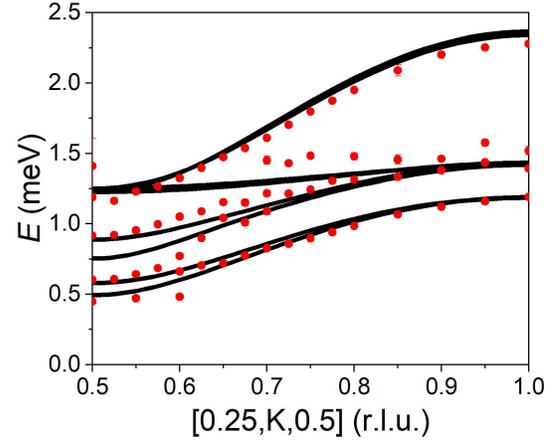}
\hspace{0.5cm} \caption{(Color online) Spin wave dispersion relationship along [0.25,K,0.5] direction, the experimental data (red points) are taken from Ref. \cite{Xiao2016}, fitted by the spin-canting model (solid line).}
\end{figure}

\begin{figure}
\centering
\includegraphics[width=8.5cm]{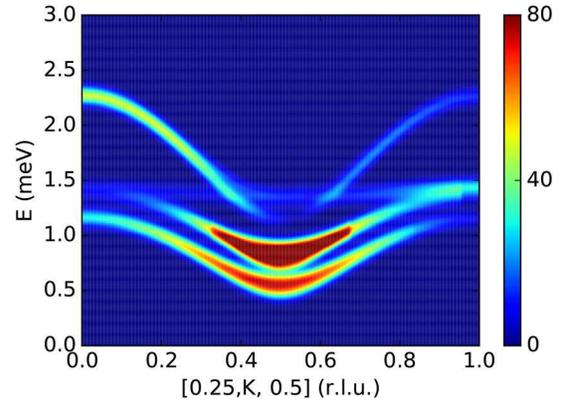}
\hspace{0.5cm} \caption{(Color online) Spin wave excitation spectrum along [0.25,K,0.5] direction through the magnetic peak (0.25,0.5,0.5).}
\end{figure}

Taking into account the spin wave dispersion along the high symmetry directions mentioned above, the neutron scattering intensity maps calculated with the two differently spin-canted magnetic moments at Mn$_a$ and Mn$_b$ are quite consistent with the observed neutron scattering spectra. Since 11 parameters are sufficient to show a good agreement with the observed data, the current study involves no further parameters dictating the magnetic coupling of Mn-Mn pairs with a longer interaction distance. However, the present model is still impossible to describe the other low-lying energy excitation mode $\omega_1$=0.07 meV observed in Ref.\cite{Xiao2016}. One promising technique to clarify the origin or the character of this excitation is polarized neutron scattering.

\section{IV. CONCLUSIONS}

In the presence of two spin-canting textures relying on a weak intrinsic polarity in the nuclear structure of MnWO$_4$, a more reliable model for theoretical spin-wave excitations could be provided for its AF1 phase. In comparison with previous inelastic neutron scattering spectra, it is shown that the spin-wave dispersions of this phase could be properly described by a Heisenberg model with 11 magnetic exchange couplings and single-ion anisotropy parameters. It is confirmed that long-range AF interactions are dominant in AF1 as our spin-wave dispersion relationship could be fitted well with all negative exchanging coupling constants. A strong variation in their magnitudes with increasing the Mn-Mn distance reflects strongly geometrically frustrated zigzag-like spin chains in AF1.

It should be noted that recent neutron scattering experiment observed two low energy excitations $\omega_1$ and $\omega_2$ with energy gaps at 0.07 meV and 0.45 meV, respectively \cite{Xiao2016}. Both of them cannot be described by the Heisenberg Hamiltonian based on the previous collinearly aligned spin configuration $\uparrow\uparrow\downarrow\downarrow$. In previous work, they were regarded as electromagnon excitations which might arise from the DM interaction. Interestingly, with the new non-collinear magnetic model, the $\omega_2$ excitation mode is properly described and we assume that it could be the lowest spin wave branch. However, our model still failed to interpret the other low-lying excitation mode $\omega_1$ with energy gap of 0.07 meV at the magnetic zone centre. Further polarized neutron scattering measurements would be helpful to understand the nature of this excitation as well as the mechanism of magnetoelectric coupling in MnWO$_4$.

\section*{ACKNOWLEDGMENTS}
B.-Q. Liu is supported by China Scholarship Council, the National Natural Science Foundation of China (No.11305150, 11674406), and Science Challenge Project (No.TZ2016004). S.-H. Park acknowledges German Federal Ministry of Education and Research (BMBF) for the financial support via 05K13WMB.


\end{document}